\documentclass[aps,pra,preprint]{revtex4-1}
\usepackage[mathscr]{euscript}
\usepackage{color,soul}
\usepackage{cancel}
\usepackage[normalem]{ulem}
\usepackage{float,epsfig}
\usepackage{bm}
\usepackage{graphicx}
\usepackage{subfigure}
\usepackage{appendix}
\usepackage{mathrsfs}
\usepackage[colorlinks=true,linkcolor=blue,urlcolor=blue]{hyperref}
\newcommand{\bea}{\begin{eqnarray}}
\newcommand{\eea}{\end{eqnarray}}
\newcommand{\beq}{\begin{equation}}
\newcommand{\eeq}{\end{equation}}

\def\/{\over}

\begin{document}

\setstcolor{red}

\title{Quantum corrections to the classical electrostatic interaction between induced dipoles}


\author{Yongshun Hu}
\email[]{ys-hu@qq.com}
\affiliation{Department of Physics and Synergetic Innovation Center for Quantum Effects and Applications, Hunan Normal University, Changsha, Hunan 410081, China}
\author{Jiawei Hu}
\email[Corresponding author. ]{jwhu@hunnu.edu.cn}
\affiliation{Department of Physics and Synergetic Innovation Center for Quantum Effects and Applications, Hunan Normal University, Changsha, Hunan 410081, China}
\author{Hongwei Yu}
\email[Corresponding author. ]{hwyu@hunnu.edu.cn}
\affiliation{Department of Physics and Synergetic Innovation Center for Quantum Effects and Applications, Hunan Normal University, Changsha, Hunan 410081, China}



\begin{abstract}
We study, in the presence of an external electrostatic field, the interatomic  interaction between two ground-state atoms coupled with vacuum electromagnetic fluctuations within the dipole coupling approximation
based on the perturbation theory. We show that,  up to the fourth order, the electrostatic-field-induced interatomic interaction is just the classical dipole-dipole interaction, which disagrees with the recent result from Fiscelli {\it et al.} [G. Fiscelli {\it et al.}, Phys. Rev. Lett. \textbf{124}, 013604 (2020)].
However, to higher orders, there exist external-field-related
quantum corrections to the induced classical electrostatic dipole-dipole interaction.
In the sixth order,  the external field effectively modifies the atomic polarizability to give rise to a two-photon-exchange quantum correction, while in the eighth order, the external field enables an additional process of three-photon exchange which is not allowed in the absence of  the external field, and this process generates an $r^{-11}$ term in the interaction potential in the far regime, where $r$ is the interatomic separation.
Numerical estimations show that these external-field-related quantum corrections are  much smaller than the two-photon-exchange Casimir-Polder interaction.

\end{abstract}


\maketitle


\section{Introduction}
\setcounter{equation}{0}

In a quantum sense, there inevitably exist quantum vacuum fluctuations, which may induce some novel effects that do not exist in classical physics. A typical example is the Casimir-Polder (CP) interaction \cite{CP}.
In general, fluctuating electromagnetic fields in vacuum induce instantaneous electric dipole moments in neutral atoms and molecules, which then interact with each other via the exchange of virtual photons.
Such effects  have been widely applied in various areas, e.g., in material science, chemistry, and biology
(see Ref. \cite{Woods2016} for a recent review).
Moreover,  the measurement of the atom-surface and surface-surface CP interactions has already been accomplished \cite{Sukenik1993,Chen2006,Hertlein2008,Bender2010}. However, the interatomic CP interaction is so weak that it has  not yet been measured.

It has been shown that the interatomic CP interaction can be significantly modified by  externally applied electromagnetic waves \cite{Thirunamachandran1980,Milonni1992,Milonni1996,Bradshaw2005,Andrews2006,Salam2006,Salam2007, Sherkunov2009}.
Compared to the interatomic CP interaction, which originates purely from the electromagnetic vacuum fluctuations, the physical process resulting in the leading interatomic interaction in the presence of external electromagnetic waves occurs differently. In the latter case, the externally applied electromagnetic waves induce dipole moments in atoms, which are then correlated to each other through the electromagnetic vacuum fluctuations. That is, a real photon is scattered by a pair of atoms which are coupled via the exchange of a virtual photon, and the interaction energy is thus modified. Such a modified CP interaction, which is actually a fourth-order effect, has been shown to behave as $r^{-3}$ in the near regime and $r^{-2}$ or $r^{-1}$ in the far regime depending on the propagating direction of the external  electromagnetic field with respect to the alignment of the atoms, where $r$ is the interatomic distance \cite{Thirunamachandran1980}. Moreover, this result reduces to the classical dipole-dipole interaction, which behaves as $r^{-3}$ in all distance regimes, when the frequency of the external electromagnetic field tends to zero. Naturally, one may wonder what the interatomic interaction will be in the presence of  externally applied electrostatic fields.

First of all, external electrostatic fields will induce electric  dipole moments in atoms  and so
a classical dipole-dipole interaction will result. However, to higher orders, there exist quantum corrections that arise from modified vacuum field fluctuations due to the presence of the external fields.
In fact, the interaction between two atoms or molecules in the presence of an electrostatic field has been studied by Mackrodt \cite{Mackrodt1974}.  It has been shown that, in the fourth order, the interatomic interaction contains two parts:  One is the classical dipole-dipole interaction induced by the electrostatic field, which behaves as $r^{-3}$ in all distance regimes and can be obtained directly through the fourth-order perturbation theory with the assistance of 24 one-photon-exchange time-ordered diagrams, while the other is the well-known two-photon-exchange CP interaction induced purely by vacuum fluctuations.
Recently, this question was revisited by Fiscelli \textit{et al.} in Ref. \cite{Passante2020} in a different approach, wherein the authors  first obtained the dressed  atomic states in the electrostatic field and then calculated the second-order energy correction between dressed atoms due to the coupling with the fluctuating electromagnetic fields. They found that, in the fourth order,  the  interatomic interaction from one-photon exchange can generate a novel $r^{-4}$  term in the far zone.
That is, there are different answers to the question of whether the  induced classical electrostatic dipole-dipole interaction can be  corrected by electromagnetic vacuum fluctuations or, in other words, whether external-field-related quantum corrections to the classical electrostatic dipole-dipole interaction exist besides those originating purely from quantum  vacuum fluctuations.

In this paper we consider, in the presence of an external electrostatic field, the interatomic  interaction in the fourth, sixth and eighth orders within the dipole coupling approximation, based on the perturbation method utilized in Ref. \cite{Passante2020}.
First, we give a formal analysis of the  interaction between two ground-state atoms in the presence of an electrostatic field.
As we will show in detail, in the fourth order, the external electrostatic field merely induces a classical  dipole-dipole interaction between the two atoms and does not generate an $r^{-4}$ term in the far zone, contrary to the recent claim in Ref. \cite{Passante2020}. In the sixth order, the  external electrostatic field  effectively modifies the atomic polarizability to affect the two-photon-exchange interatomic CP interaction,
which can  be considered as  one kind of external-field-related quantum correction to the classical dipole-dipole interaction,  while in the eighth order, the external field enables an additional three-photon exchange which does not occur between two ground-state atoms in vacuum, which is another kind of external-field-related quantum correction.
Second, as a specific example, we apply our analysis
to the case of two ground-state hydrogen atoms and show concretely what the electrostatic-field-related quantum corrections of the induced classical dipole-dipole interaction are and whether such corrections may be detectable in the foreseeable future.
Throughout this paper, the Einstein summation convention for repeated indices is assumed and the Latin indices run from $1$ to $3$.

\section{Interatomic interaction in the presence of an electrostatic field}
\label{CPEF}
We now consider the interatomic interaction  between two neutral atoms ($A$ and $B$) in their ground states in the presence of an electrostatic field based on the perturbation method. 
Since our concern is the  interatomic interaction related to the electrostatic field, we first obtain the atomic states dressed by the external electrostatic field to the second order and then calculate the second-order, fourth-order, and sixth-order energy corrections of the dressed ground state, due to the interaction with the fluctuating transverse fields.

The total Hamiltonian of the system considered is given by
\beq \label{Hamiltonian}
H=H_F+H_S+V_S+H_I,
\eeq
where $H_{F}$ and $H_{S}$ are the Hamiltonians of the fluctuating vacuum electromagnetic fields and the atoms, respectively, and $V_S$ and $H_I$ are respectively the interaction Hamiltonians of the atoms with the external electric field and the fluctuating vacuum electromagnetic fields, which, in the multipolar coupling scheme and within the dipole approximation, take the form
\bea
V_S &=& -d^A_i \varepsilon_{i}(\vec x_A)-d^B_i \varepsilon_{i}(\vec x_B),\\
H_I &=& -\epsilon_0^{-1}d^A_i \mathscr{E}_i(\vec x_A)-\epsilon_0^{-1}d^B_i \mathscr{E}_i(\vec x_B).
\eea
Here $d^{\sigma}_i$ is the component of the dipole moment of atom $\sigma$, $\epsilon_0$ is the vacuum permittivity, $\varepsilon_i(\vec x)$ represents the external electrostatic field, and  $\mathscr{E}_i(\vec x)$ is the transverse displacement field, of which the explicit form  is given by \cite{PT,Passante2018}
\beq\label{E-field}
\mathscr{E}_i(\vec x)=i\sum_{\vec k, \lambda}\sqrt{\frac{\hbar c k \epsilon_0}{2\mathcal{V}}} e^{\lambda}_i [a_{\lambda}(\vec k) e^{i \vec k \cdot \vec x}-a^{\dag}_{\lambda}(\vec k)e^{-i \vec k \cdot \vec x}],
\eeq
where $\mathcal{V}$ is the quantization volume, $a_{\lambda}(\vec k)$ and $a^{\dag}_{\lambda}(\vec k)$ are respectively the annihilation and creation operators of the electromagnetic vacuum field with wave vector $\vec k$ and polarization $\lambda$,  and $e^{\lambda}_i$ are polarization unit vectors. 

We label the unperturbed energy eigenvalues of the ground and excited states as $E_0$ and $E_{n}$ $( n=1,2,3, . . .)$, respectively, with the corresponding eigenstates  being  $|e_0\rangle$ and $|e_n\rangle$.
In the presence of an electrostatic field, the states of neutral atoms are modified due to the interaction between the induced permanent dipole moments and the electrostatic field. Formally, we denote the corrected atomic state by $|\hat e_{\gamma}\rangle$, which can be given by the second-order perturbation theory, i.e.,
\bea\label{eq2}
\nonumber|\hat e_{\gamma} \rangle&=&\left(1-\sum_{\alpha} {'}\frac{\hat d^{\gamma \alpha}_i \hat d^{\alpha \gamma}_j} {2E_{\gamma\alpha}^2}\varepsilon_i \varepsilon_j \right)|e_{\gamma}\rangle -\sum_{\alpha}{'}\frac{\hat d^{\alpha \gamma}_i \varepsilon_i }{E_{\gamma\alpha}}|e_{\alpha}\rangle\\
&&+\sum_{\beta} {'}\left(\sum_{\alpha} {'}\frac{\hat d^{\beta\alpha}_i \hat d^{\alpha \gamma}_j } {E_{\gamma\alpha}E_{\gamma\beta}}-\frac{\hat d^{\beta \gamma}_i \hat d^{\gamma\gamma}_j }{E_{\gamma\beta}^2}\right)\varepsilon_i\varepsilon_j |e_{\beta}\rangle,
\eea
where $\alpha, \beta, \gamma=0, 1, 2, 3, . . .$, 
$E_{\gamma\alpha}=E_{\gamma}-E_{\alpha}$, and $\hat d^{\alpha\beta}_i=\langle e_{\alpha}|d_i|e_{\beta}\rangle$. 
Note here that a neutral atom becomes a polar one in the presence of the electrostatic field.

First, we calculate the interaction due to the exchange of a single virtual photon between the two atoms dressed by the electrostatic field, i.e., the fourth-order effect.
Let us note that the one-photon exchange  is only possible for dressed atoms since the dressed  atoms are in a superposition of the ground state and excited states [see Eq. (\ref{eq2})], while the undressed ones are in the energy eigenstates, and the average value of the induced dipole moment over dressed atomic states is nonzero, while it is zero over undressed atomic states.
Based on the  second-order perturbation theory, we obtain \cite{PT}
\beq\label{E_AB perturbation}
\Delta E_{AB}=\sum_{I}\frac{\langle \psi|H_I|I\rangle\langle I|H_I|\psi\rangle}{E_{\psi} -E_{I}},
\eeq
where $|\psi\rangle=|\hat e^A_0\rangle |\hat e^B_0\rangle|0\rangle$ is the corrected initial state of the system, with $|0\rangle$ the electromagnetic vacuum state, $|I\rangle=|\hat e^A_0\rangle |\hat e^B_0\rangle |1_{\vec k, \lambda}\rangle$ is the virtual intermediate state, and $E_I$ is the energy of state $|I\rangle$. Then the interaction energy can be obtained as
\beq
\Delta E_{AB}=-\sum_{\vec k,\lambda}\frac{1}{\epsilon_0 V} \hat d^A_i \hat d^B_j  e^{\lambda}_i e^{\lambda}_j e^{i\vec k\cdot \vec r},
\eeq
where $\vec r=\vec x_A-\vec x_B$  and $\hat d^{\sigma}_i=\langle\hat e_0^{\sigma}|d^{\sigma}_i|\hat e_0^{\sigma}\rangle$ is the induced permanent dipole moment in atom $\sigma$. Replacing  summation by  integration $\sum_{\vec k}\rightarrow V/(2\pi)^3 \int k^2 dk \int d\Omega$ and performing the polarization summation $\sum_{\lambda}e^{\lambda}_ie^{\lambda}_j=\delta_{ij}-\hat k_i \hat k_j$, we obtain
\beq\label{classical dd}
\Delta E_{AB}=\frac{\hat d^A_i \hat d^B_j}{4\pi\epsilon_0 r^3}(\delta_{ij}-3\hat r_i \hat r_j),
\eeq
where $\hat r_i$ is the $i$th component of the unit vector $\vec r /r$. That is, in the fourth-order, the electrostatic-field-induced interatomic interaction is just the $r^{-3}$ classical dipole-dipole interaction which applies in all distance regimes. Notice that here the result is obtained by a second-order calculation of two atoms dressed by an electrostatic field to the second order, which is in agreement with an early result between two undressed atoms obtained by a fourth-order calculation \cite{Mackrodt1974}.  Moreover, our result is  also in agreement with the interatomic interaction in the presence of external electromagnetic waves  \cite{Thirunamachandran1980,Milonni1996} when the wave frequency is so small that the field is effectively static and with the electrostatic interaction between two atomic dipole moments induced by an external static field in a cavity \cite{Donaire3} when the thickness of the cavity tends to infinity. However, the above result does not agree with that in Ref. \cite{Passante2020}.
In Ref. \cite{Passante2020} the ground state was corrected to the second order while the excited state was only corrected to the zeroth order, so the corrected excited states used in Ref. \cite{Passante2020} were not orthogonal to the corrected ground state (under the second order). Then a nonzero incorrect $r^{-4}$ term in the far regime was obtained since the authors chose the
excited states corrected to the zeroth order    as the intermediate states. Actually, the contributing term is not relevant to the corrected excited states if the excited states are also corrected to the second order, due to the fact that they are orthogonal  to the
corrected ground state in this circumstance.

Second, we examine the interaction between the two dressed atoms due to  two-photon exchange. For simplicity, we consider only the far-zone behavior of the dipole-dipole interaction potential. In the far zone, the computation  relates to only four time-ordered diagrams \cite{PT,Salam} and the expression of the interaction energy shift is formally  given by
\bea\label{Integral_2FZ}
\nonumber\Delta E_{AB}&=&-\sum_{\vec k, \vec k'}\left(\frac{\hbar c k}{2\epsilon_0 \mathcal{V}}\right) \left(\frac{\hbar c k'}{2\epsilon_0 \mathcal{V}}\right)\frac{e^{i (\vec k+\vec k')\cdot \vec r}}{\hbar c k+\hbar c k'} \\
&&\times(\delta_{il}-\hat k_i \hat k_l) (\delta_{jm}-\hat k_j' \hat k_m') \alpha^A_{ij}(0)\alpha^B_{lm}(0) .
\eea
Here $\alpha_{ij}(0)$ is the static polarizability, which is  defined as
\beq\label{alpha}
\alpha_{ij}(0)=\sum_{s}\frac{1}{E_{s0}}(\hat d^{0s}_{i}\hat d^{s0}_j+\hat d^{0s}_{j}\hat d^{s0}_i),
\eeq
where the index $0$ represents the corrected ground state $|\hat e_0\rangle$  and the index $s$ represents the corrected excited state $|\hat e_s\rangle$. Replacing the summation in Eq. (\ref{Integral_2FZ}) by  integration and performing the integral with the help of the integral representation
\beq
\frac{1}{k+k'}=r \int_{0}^{\infty} e^{-(k+k')r\eta}d\eta,
\eeq
we obtain
\beq\label{E_AB-2FZ}
\Delta E_{AB}=-\frac{\hbar c}{128\pi^3\epsilon_0^2r^7}M_{AB},
\eeq
where
\bea\label{M_AB}
\nonumber M_{AB}&=&20\alpha^A_{33}(0)\alpha^B_{33}(0)+13\alpha^A_{11}(0)\alpha^B_{11}(0)+13\alpha^A_{22}(0)\alpha^B_{22}(0)\\
\nonumber&&+13\alpha^A_{12}(0)\alpha^B_{12}(0)+13\alpha^A_{21}(0)\alpha^B_{21}(0)-15\alpha^A_{13}(0)\alpha^B_{13}(0)\\
&&-15\alpha^A_{31}(0)\alpha^B_{31}(0)-15\alpha^A_{23}(0)\alpha^B_{23}(0)-15\alpha^A_{32}(0)\alpha^B_{32}(0).
\eea
The above result (\ref{E_AB-2FZ}) is formally the same as the  interatomic CP interaction between undressed atoms, while the atomic polarizability is now modified by the applied electrostatic field, i.e., the elements of the dipole transition matrix in Eq. (\ref{alpha}) are related to the dressed atomic states. That is, in the sixth order,
the external-field-related quantum correction to the classical dipole-dipole interaction takes the same form  as the CP interaction but the atomic polarizability is effectively dressed by the external electrostatic field. 
Note here that the above result (\ref{E_AB-2FZ}) is in fact the far-zone counterpart of the near-zone London interaction between a pair of undressed molecules in an electrostatic field  studied in  Ref. \cite{Mackrodt1974},
which is not the complete sixth-order potential. In a complete sixth-order calculation of the interaction energy shift between two undressed molecules \cite{Mackrodt1974}, there also exist terms corresponding to a vacuum-fluctuation-induced modification of the two-photon-exchange CP interaction, i.e., an effective dressing of the atomic polarizability by vacuum fluctuations.
We leave an order-of-magnitude estimation of such an effect to Sec. \ref{3}, since here our main concern is the  interatomic  interaction relevant to the external electrostatic field.

Third,  we study the  interaction due to  the three-photon exchange  between the two dressed atoms. This is particularly interesting since the three-photon exchange is not allowed without the presence of an external electrostatic field for a reason we will explain later.   A complete calculation of the three-photon-exchange interatomic interaction involves $360$ time-ordered diagrams.
However, in the far-zone limit, the leading term  relates only to $12$  time-ordered diagrams \cite{Salam1997} and can be formally expressed as
\bea\label{Integral_3FZ}
\nonumber\Delta E_{AB}=&-&\frac{1}{3}\sum_{\vec k, \vec k', \vec k''}
\left(\frac{\hbar c k}{2\epsilon_0 \mathcal{V}}\right)\left(\frac{\hbar c k'}{2\epsilon_0 \mathcal{V}}\right) \left(\frac{\hbar c k''}{2\epsilon_0 \mathcal{V}}\right)\\
\nonumber&\times& (\delta_{il}-\hat k_i \hat k_l)(\delta_{jm}-\hat k_j' \hat k_m')(\delta_{kn}-\hat k_k'' \hat k_n'')\\
&\times&\frac{e^{i(\vec k+\vec k'+\vec k'') \cdot \vec r}}{\hbar c(k+k'+k'')}\beta^A_{ijk}(0) \beta^B_{lmn}(0),
\eea
where $\beta_{ijk}(0)$ is the static first hyperpolarizability tensor, which is formally given by
\bea\label{betaijk}
\nonumber\beta_{ijk}(0)&=&\sum_{t,s}\frac{1}{E_{t0}E_{s0}}(\hat d^{0 t}_i \hat d^{t s}_j \hat d^{s 0}_k +\hat d^{0 t}_i \hat d^{t s}_k \hat d^{s 0}_j+\hat d^{0 t}_j \hat d^{t s}_k \hat d^{s 0}_i \\
&&\quad +\hat d^{0 t}_j \hat d^{t s}_i \hat d^{s 0}_k+\hat d^{0 t}_k \hat d^{t s}_i \hat d^{s 0}_j+\hat d^{0 t}_k \hat d^{t s}_j \hat d^{s 0}_i),
\eea
where the index $0$ represents the corrected ground state $|\hat e_0\rangle$, while the indices $t$ and $s$ represent the corrected excited state $|\hat e_t\rangle$ and $|\hat e_s\rangle$, respectively.
Replacing the summation in Eq. (\ref{Integral_3FZ}) by a spherical coordinate integral and performing the integral with the help of the integral representation
\beq
\frac{1}{\omega+\omega'+\omega''}=r \int_{0}^{\infty} e^{-(\omega+\omega'+\omega'')r\eta}d\eta,
\eeq
we obtain
\beq\label{E_AB-3FZ}
\Delta E_{AB}=\frac{\hbar^2 c^2}{2^{13}\pi^5 \epsilon_0^3 r^{11}}D_{AB},
\eeq
where
\bea\label{D_AB}
\nonumber D_{AB}&=&-336\beta^A_{333}(0)\beta^B_{333}(0)+225\beta^A_{222}(0)\beta^B_{222}(0)\\
\nonumber&&+225\beta^A_{111}(0)\beta^B_{111}(0)+675\beta^A_{212}(0)\beta^B_{212}(0)\\
\nonumber&&+675\beta^A_{121}(0)\beta^B_{121}(0)-744\beta^A_{131}(0)\beta^B_{131}(0)\\
\nonumber&&-744\beta^A_{232}(0)\beta^B_{232}(0)+840\beta^A_{313}(0)\beta^B_{313}(0)\\
&&+840\beta^A_{323}(0)\beta^B_{323}(0)-1488\beta^A_{123}(0)\beta^B_{123}(0).
\eea
The above result (\ref{E_AB-3FZ}) shows that the interatomic interaction potential between the two dressed atoms due to the exchange of three virtual photons behaves as $r^{-11}$ in the far zone, which is another kind of quantum correction to the classical electrostatic dipole-dipole interaction.
In fact, the three-photon-exchange process is not possible for undressed atoms within the dipole coupling approximation, since the  undressed atoms  are in their energy eigenstates. According to the dipole transition rule, the hyperpolarizability  (\ref{betaijk})  for undressed atoms must be zero since the three  dipole transition matrix elements in the products in Eq. (\ref{betaijk}) cannot be nonzero at the same time. That is, the undressed atoms cannot return to the initial  state as required by the perturbation approach  after an exchange of odd virtual photons, since the angular momentum quantum number must be changed due  to the dipole transition rule. However, there is no such problem for dressed atoms, since  the dressed  atoms are in a superposition of the ground state and excited states [see Eq. (\ref{eq2})].

Therefore,  the possible electrostatic-field-related quantum corrections to the induced classical dipole-dipole interaction between two ground-state atoms can be classified into two categories: (i) corrections arising from the exchange of even virtual photons, in which the atomic polarizability is effectively modified by the external fields, and (ii) corrections arising from the exchange of odd virtual photons enabled by electrostatic fields, which do not occur within the dipole coupling approximation in the absence of an external electrostatic field.
Note that the atomic polarizability can also be modified by electromagnetic vacuum fluctuations  \cite{Bullough1,Bullough2,Bullough3,Bullough4,Bullough5,Donaire1,Donaire2}. Since
our main concern here is the interatomic interaction relevant to the external electrostatic field,
an order-of-magnitude estimation of such an effect is left to Sec. \ref{3}.

\section{Interaction between ground-state hydrogen atoms in an electrostatic field}\label{3}

Now we take hydrogen atoms as an example to show the  electrostatic-field-related  quantum corrections of the induced classical dipole-dipole interaction.
The unperturbed ground state of the total system is
\beq
|\phi^A_{100}\rangle|\phi^B_{100}\rangle|0\rangle,
\eeq
where $|\phi_{nlm}\rangle$ is the atomic state with quantum numbers $n$, $l$, and $m$ and energy $E_n$, and $|0\rangle$ is the electromagnetic vacuum state.
For simplicity, we assume that the direction of the external electrostatic field is along the $z$ axis, i.e., $\varepsilon_i(\vec x_{\sigma})=(0, 0, \varepsilon_{\sigma})$, and consider  the contributions from  intermediate states with $n=2$ only. For atom $A$ (or $B$), the corrected ground state is obtained as
\bea
\nonumber|\psi^A_1\rangle&=&\left(1-\gamma^2 \varepsilon_A^2 \right)|\phi_{100}\rangle-\sqrt{2}\gamma \varepsilon_A |\phi_{210}\rangle\\
&&-\frac{3^6}{2^6\sqrt{2}} \gamma^2 \varepsilon_A^2 |\phi_{200}\rangle,\label{HydrogenA-ground}
\eea
where $\gamma=2^9 q a_0/3^6 E_1$ with $q$ the electric charge and $a_0$ the Bohr radius, and $E_2=E_1/4$ has been applied. Then the second-order ground state of the system (eigenstate of $H_F+H_S+V_S$) is given by
\beq\label{Hydrogen-ground}
|\psi\rangle=|\psi^A_1\rangle|\psi^B_1\rangle|0\rangle.
\eeq
To calculate the  interaction energy in the presence of the external electrostatic field, we need to find all possible intermediate states.
For this purpose, we should first  find  the corrected  excited states of the atoms in the presence of the electrostatic field to the second order.
For a hydrogen atom, the unperturbed fourfold-degenerate excited state ($n=2$) is divided into three energy levels (one degenerate and two nondegenerate) in the presence of an external electrostatic field, i.e., the Stark effect. 
Therefore, the second-order correction of the two nondegenerate levels can be approximately obtained by nondegenerate-state perturbation theory. For atom $A$ (or $B$), the corrected excited states can be obtained as (consider only the contributions from $n=1$ and $2$)
\bea
\nonumber |\psi^A_{21}\rangle&=&\frac{1}{\sqrt{2}}\left(1-\frac{1}{2}\gamma^2\varepsilon_A^2\right) \left(|\phi^A_{200}\rangle-|\phi^A_{210}\rangle\right)\\
&&-\left(\gamma\varepsilon_A- \frac{3^6}{2^7}\gamma^2\varepsilon_A^2 \right)|\phi^A_{100}\rangle, \label{Hydrogen-excited1}\\
\nonumber |\psi^A_{22}\rangle&=&\frac{1}{\sqrt{2}}\left(1-\frac{1}{2}\gamma^2\varepsilon_A^2\right) \left(|\phi^A_{200}\rangle+|\phi^A_{210}\rangle\right) \\
&&+\left(\gamma\varepsilon_A+ \frac{3^6}{2^7}\gamma^2\varepsilon_A^2 \right)|\phi^A_{100}\rangle.\label{Hydrogen-excited2}
\eea
Obviously, the corrected excited states (\ref{Hydrogen-excited1}) and (\ref{Hydrogen-excited2}) are orthogonal to the ground state (\ref{HydrogenA-ground}) (in the second order, which is proportional to $q^2$), i.e., $\langle\psi^{\sigma}_{21}|\psi^{\sigma}_1\rangle=\langle\psi^{\sigma}_{22}|\psi^{\sigma}_1\rangle$ can be taken as zero in the calculations. This is also required because  the new eigenstates (ground or excited) of the atoms in the presence of an electrostatic field should still be orthogonal.

Now we  show the  electrostatic-field-related quantum corrections to the induced classical dipole-dipole interaction between two ground-state atoms.
First, we consider the correction related to the process of two-photon exchange between two dressed hydrogen atoms, wherein the atomic polarizability is  effectively modified by the electrostatic field.
Although the interaction energy shift (\ref{E_AB-2FZ}) is relative to all components of the polarizability tensor, the nonvanishing modification part is only in $\alpha_{33}(0)$ since  the external electric field is assumed to be along the $z$ axis. According to Eqs. (\ref{HydrogenA-ground}), (\ref{Hydrogen-excited1}), and (\ref{Hydrogen-excited2}), it is easy to obtain
\beq\label{alpha33}
\alpha_{33}(0)=-\frac{2^{18}}{3^{11} E_1}q^2 a_0^2-\frac{2^{22}}{3^{11}E_1}q^2a_0^2\left(\frac{q a_0 \varepsilon}{E_1}\right)^2.
\eeq
The first term is the undressed polarizability and the second term is the polarizability  modified by the electrostatic field $\varepsilon$.
The interaction potential relevant to the electrostatic field is then reflected in the first term in Eq. (\ref{M_AB}), i.e.,
\beq
\alpha^A_{33}(0)\alpha^B_{33}(0)=\frac{2^{36}}{3^{22}E_1^2}q^4 a_0^4+\frac{2^{40}}{3^{22}E_1^4}q^6 a_0^6(\varepsilon_A^2+\varepsilon_B^2) +\frac{2^{44}}{3^{22}E_0^6}q^8a_0^8\varepsilon_A^2\varepsilon_B^2.
\eeq
Thus, in the sixth order, the electrostatic-field-related quantum correction to the induced classical dipole-dipole interaction is obtained as
\beq\label{E^M}
\Delta E^{(6)}_{AB}=-5\frac{2^{35}\hbar c q^6 a_0^6}{3^{22}\pi^3\epsilon_0^2 E_1^4 r^7}(\varepsilon_A^2+\varepsilon_B^2).
\eeq
Now we compare numerically the external-field-related quantum correction with respect to the unperturbed CP potential between two ground-state atoms due to the exchange of two virtual photons  in vacuum~\cite{CP}.
Here we consider only the contributions from $n=1$ and $2$ and assume that the atoms are isotropically polarizable, i.e., $\alpha_{ij}=\alpha\, \delta_{ij}$. Then the  CP potential takes the form
\beq
U_{AB}=-\frac{23\hbar c}{64\pi^3\epsilon_0^2 r^7}\alpha^A \alpha^B =-23\frac{2^{30}\hbar c q^4 a_0^4}{3^{22}\pi^3\epsilon_0^2 E_1^2 r^7},
\eeq
where $\alpha^{A (B)}$ is the scalar polarizability of atom $A$ or $B$ defined as
\beq
\alpha=\sum_{s}\frac{2}{3E_{s0}}\hat d_{i}^{0s} \hat d_{i}^{s0}. 
\eeq
Taking $\varepsilon_A=\varepsilon_B\simeq10^8$ V/m, which is within the applicable range of the perturbation theory and can be reached in the laboratory \cite{Maron1986,Bailey1995}, and $r=10^{-6}$ m, which is larger than the transition wavelength ($\sim 10^{-8}$ m), the relative magnitude of the corrected term with respect to the unperturbed CP potential can be obtained as
\beq
\frac{\Delta E^{(6)}_{AB}}{U_{AB}}=\frac{160}{23E_1^2}q^2 a_0^2(\varepsilon_A^2 + \varepsilon_B^2)\simeq10^{-6}.
\eeq
That is, the  electrostatic-field-related quantum correction in the sixth order is smaller than the  unperturbed two-photon-exchange CP interaction.
This is expected because the former is a sixth-order effect while the latter is a fourth-order effect.


Second, we calculate the  quantum correction to the classical electrostatic dipole-dipole interaction related to the  process of three-photon exchange between two dressed hydrogen atoms.
According to Eqs. (\ref{HydrogenA-ground}), (\ref{Hydrogen-excited1}) and (\ref{Hydrogen-excited2}), it is easy to obtain the nonvanishing component of the static first hyperpolarizability,  i.e.,
\beq
\beta_{333}(0)=\frac{2^{38}}{3^{22}E_1^3}q^4 a_0^4 \varepsilon.
\eeq
Then the  interaction energy (\ref{E_AB-3FZ}) can be expressed as
\beq\label{E3_AB hydrogen}
\Delta E_{AB}^{(8)}=-7\frac{2^{67}\hbar^2 c^2 q^8 a_0^8 }{3^{43}\pi^5 \epsilon_0^3 E_1^6 r^{11}}\varepsilon_A \varepsilon_B.
\eeq
This term is absent without the electrostatic field since the   exchange of three virtual photons between two ground-state hydrogen atoms  is forbidden in the absence of an electrostatic  field.
Taking $r=10^{-6}$ m and $\varepsilon_A=\varepsilon_B\simeq10^8$ V/m, the relative magnitude of  this interaction energy with respect to the unperturbed two-photon-exchange CP potential is obtained as
\beq
\frac{\Delta E_{AB}^{(8)}}{U_{AB}}=\frac{7\times 2^{37}}{23\times3^{21}}\frac{\hbar c q^4 a_0^4} {\pi^2\epsilon_0 E_1^4 r^4}\varepsilon_A \varepsilon_B\simeq10^{-20}.
\eeq
That is, the electrostatic-field-enabled three-photon-exchange interatomic interaction is much smaller than the  unperturbed 
two-photon-exchange CP interaction, which suggests that an experimental observation of such an effect seems to be a hope for the distant future.

Finally, we note that, apart from the electrostatic-field-related quantum correction to the induced classical dipole-dipole interaction between two ground-state atoms we considered here, there are other terms in the   complete result of the interatomic interaction. Now we would like to make an order-of-magnitude estimation of these  terms and compare them with the  quantum corrections of the classical electrostatic dipole-dipole interaction given in Eqs.  (\ref{E^M}) and (\ref{E3_AB hydrogen}).


(i) As mentioned in Sec. \ref{CPEF}, in the sixth order, the two-photon-exchange  CP interaction can also be modified by vacuum fluctuations, which is relevant to the self-interaction processes and can be  regarded as causing an effective modification of the denominators of the atomic polarizability due to the vacuum-induced energy shift \cite{Bullough1,Bullough2,Bullough3,Bullough4,Bullough5,Donaire1,Donaire2}.
As an order-of-magnitude estimation, we write the vacuum-modified atomic polarizability
 in analogy to the electrostatic-field-modified polarizability (\ref{alpha33}) as
\beq
\alpha\sim-\frac{q^2 a_0^2}{E_1}-\frac{q^2 a_0^2}{E_1}\left(\frac{\hbar\delta\omega_L}{E_1}\right)^2.
\eeq
Here $\hbar\delta\omega_L$ is the vacuum-induced Lamb shift, which plays a  role similar to that of the  electrostatic-field-induced Stark shift ($\sim q a_0 \varepsilon$).
Then the leading modification of the two-photon-exchange CP potential induced by vacuum fluctuations (in the sixth-order) can be approximately given as
\beq\label{E^M_L}
\Delta E^{(vac)}_{AB}\sim-\frac{\hbar c q^4 a_0^4}{\epsilon_0^2 E_1^2 r^7} \left(\frac{\hbar\delta\omega_L}{E_1}\right)^2.
\eeq
The ratio of the  result (\ref{E^M_L}) to the  sixth-order quantum correction to the classical electrostatic dipole-dipole interaction (\ref{E^M}) can be written as
\beq
\frac{\Delta E^{(vac)}_{AB}}{\Delta E^{(6)}_{AB}}\sim\frac{\hbar^2 \delta\omega_L^2}{q^2 a_0^2 \varepsilon^2}.
\eeq
As a numerical estimation, we take the Lamb shift $\hbar\delta\omega_L$  to be of the order of $\sim10^{-6}$ eV, i.e., the energy difference between $2^{2}S_{1/2}$ and $2^{2}P_{1/2}$ of a hydrogen atom \cite{Lamb1947}. To achieve  $\Delta E^{(6)}_{AB}\gg\Delta E^{(vac)}_{AB}$, it is required that
\beq
\varepsilon\gg\frac{\hbar \delta\omega_L}{q a_0}\simeq10^{4}~\text{V/m}.
\eeq
That is, when the external electrostatic field is much larger than $10^{4}~\text{V/m}$, the Stark shift is much larger than the Lamb shift and the  electrostatic-field-related quantum correction (sixth order) to the classical electrostatic dipole-dipole interaction is much larger than  the modification of the  two-photon-exchange CP interaction by vacuum fluctuations (sixth order).

(ii) As discussed in Sec. \ref{CPEF}, the  electrostatic field enables a process of three-photon exchange  for neutral atoms within the dipole  coupling approximation.
However, when the quadrupole coupling is taken into account, the three-photon-exchange processes  may also occur  without the presence of an external electrostatic field, which is related to the processes of one quadrupole  and two dipole transitions, and the corresponding  interaction potential  can be  given by the dimensional analysis as
\beq\label{E^{QD}}
\Delta E^{Q}_{AB}\sim-\frac{\hbar^2 c^2 q^6 a_0^8}{\epsilon_0^3 E_1^4 r^{13}}.
\eeq
Since it scales as $r^{-13}$, it is  usually smaller than the electrostatic-field-enabled three-photon-exchange interatomic interaction $\Delta E^{(8)}_{AB}$ in the far region.

(iii) Another point of consideration is the higher-order far-zone expansion of the two-photon-exchange CP interaction. In fact, it is well known that the $r^{-7}$  CP  potential is the leading term of the complete two-photon-exchange interatomic CP potential in the far zone, which takes the form \cite{CP}
\beq
E_{\text{CP}}=-\frac{\hbar c \alpha^A \alpha^B}{16\pi^3\epsilon_0^2 r^6}\int^{\infty}_{0}\frac{k_A^2 k_B^2 e^{-2u r}} {(k_A^2+u^2)(k_B^2+u^2)}(u^4 r^4+ 2u^3 r^3+ 5u^2 r^2+ 6u r+ 3) du,
\eeq
where $k_{A(B)}=\omega_{A(B)}/c$. For identical atoms ($k_A=k_B$), after some algebra, we obtain the far-zone expansion of the CP potential, i.e.,
\beq\label{E_CP}
E_{\text{CP}}=-\frac{23\hbar c \alpha^A \alpha^B}{64\pi^3\epsilon_0^2 r^7} \left(1-\frac{129}{23k_A^2 r^2} +\frac{1917}{23k_A^4r^4}- . . .\right).
\eeq
This shows that there  exist terms scaling as $r^{-9}$ and $r^{-11}$ in the far-zone expansion (labeled as $\Delta_1 E_{\text{CP}}$ and $\Delta_2 E_{\text{CP}}$, respectively). Since the electrostatic-field-enabled three-photon-exchange interatomic interaction $\Delta E^{(8)}_{AB}$  behaves as $r^{-11}$, it is much smaller than the $r^{-9}$ term in Eq. (\ref{E_CP}) in the far region, i.e., $\Delta E^{(8)}_{AB}\ll\Delta_1 E_{\text{CP}}$. Moreover, the relative order of magnitude between the $r^{-11}$ term (i.e., $\Delta_2 E_{\text{CP}}$) in Eq. (\ref{E_CP}) and $\Delta E^{(8)}_{AB}$ can be obtained as
\beq
\frac{\Delta_2 E_{\text{CP}}}{\Delta E_{AB}^{(8)}}\sim\frac{\epsilon_0 E_1^4 \lambda^4}{\hbar c q^4 a_0^4 \varepsilon^2},
\eeq
where $\lambda=k_A^{-1}$ is the transition wavelength.
A numerical estimation shows that the electrostatic-field-enabled three-photon-exchange interaction $\Delta E^{(8)}_{AB}$ is smaller than $\Delta_2 E_{\text{CP}}$ for realistic values of the electrostatic field. This is understandable  since the former is an eighth-order effect while the latter is the fourth-order effect.

\section{Summary}
\label{sec_disc}
In this paper we investigated the interatomic  interaction between two ground-state atoms in the presence of an external electrostatic field within the dipole coupling approximation.  Based on the perturbation method,
we demonstrated that, in the fourth order, the external electrostatic field only induces a classical dipole-dipole interaction between the two atoms, in contrast to a recent claim in Ref. \cite{Passante2020}, while in higher orders,  such a classical effect can be  corrected by two kinds of mechanisms.
In the sixth order,  the external field effectively modifies the atomic polarizability to give rise to a two-photon-exchange quantum correction, while in the eighth order,  the external field enables an additional process of three-photon exchange which is not allowed in the absence of  the external field, and this process generates an $r^{-11}$ term in the interaction potential in the far regime.
Numerical estimations showed that these external-field-related quantum corrections are  much smaller than the two-photon-exchange Casimir-Polder interaction.

\begin{acknowledgments}

This work was supported in part by the NSFC under Grants No. 11690034, No. 11805063, and No. 12075084 and the Hunan Provincial Natural Science Foundation of China under Grant No. 2020JJ3026.

\end{acknowledgments}




\end{document}